\documentclass[%
 aip,
 jmp,%
 amsmath,amssymb,
 reprint,%
 onecolumn,
unsortedaddress
]{revtex4-1}

\usepackage{graphicx}
\usepackage{dcolumn}
\usepackage{bm}

\usepackage{tikz}

\def\R{{\mathbb R}}
\def\C{{\mathbb C}}

\def\E{{\mathbb{E}}}            

\def\bx{{\mathbf{x}}}
\def\by{{\mathbf{y}}}
\def\bk{{\mathbf{k}}}
\def\bxi{{\bm{\xi}}}
\def\bphi{{\bm \phi}}
\def\bpsi{{\bm \psi}}

\def\bnu{{\bm \nu}}
\def\bp{{\mathbf{p}}}
\def\bq{{\mathbf{q}}}
\def\bh{{\mathbf{h}}}

\def\Id{\mathbf{I}}
\def\eps{\varepsilon}

\def\Tr{{\mathrm{Tr}}}

\begin{document}

\title{Radiative transfer and diffusion limits for wave field correlations in locally shifted random media}
\thanks{This work was supported  by ERC Advanced Grant Project MULTIMOD--267184.}

\author{Habib Ammari}
\email{habib.ammari@ens.fr} 
\affiliation{Department of Mathematics and Applications,
Ecole Normale Sup\'erieure, 45 Rue d'Ulm, 75005 Paris, France}

\author{Emmanuel Bossy}
\email{emmanuel.bossy@espci.fr} 
\affiliation{Institut Langevin, ESPCI ParisTech,
CNRS UMR 7587, 10 rue Vauquelin, 75231 Paris Cedex 05, France}

\author{Josselin Garnier}
\email{garnier@math.univ-paris-diderot.fr}
\affiliation{Laboratoire de Probabilit\'es et Mod\`eles Al\'eatoires \& Laboratoire Jacques-Louis Lions,
Universit\'e Paris VII, 75205 Paris Cedex 13, France}

\author{Wenjia Jing}
\email{wjing@dma.ens.fr} 
\affiliation{Department of Mathematics and Applications,
Ecole Normale Sup\'erieure, 45 Rue d'Ulm, 75005 Paris, France}

\author{Laurent Seppecher}
\email{laurent.seppecher@ens.fr} 
\affiliation{Department of Mathematics and Applications,
Ecole Normale Sup\'erieure, 45 Rue d'Ulm, 75005 Paris, France}

\date{\today}

\begin{abstract}
The aim of this paper is to develop a mathematical framework for
opto-elastography. In opto-elastography, a mechanical perturbation
of the medium produces a decorrelation of optical speckle patterns
due to the displacements of optical scatterers. To model this, we
 consider two optically random media, with the second medium obtained by shifting 
the first medium in some local region. We derive the radiative transfer
 equation for the cross-correlation of the wave fields in the  media. 
 Then we derive its  diffusion approximation. In both the
 radiative transfer and the diffusion regimes, we relate the correlation of speckle patterns
to the solutions of the radiative transfer and the diffusion equations. 
We present numerical simulations based on our model which are in agreement with
recent experimental measurements.
\end{abstract}







\maketitle

\section{Introduction}
\label{sec:intro}

When a strongly scattering medium is illuminated by a coherent laser beam, the spatial profile of the transmitted light
forms a speckle pattern
which results from the interference of many multiply scattered waves, having different phases and amplitudes.
The intensity of the speckle pattern varies randomly in space and its statistical properties depend on the 
statistical properties of the scattering medium. When the optical scatterers are displaced in some region,
the speckle pattern changes and its correlation with the original speckle pattern depends
on the amplitude of the displacement and on the local optical properties (scattering and absorption)
of the region affected by the displacement.

Focused ultrasound introduces such displacements of scatterers through two different
mechanisms. First, ultrasound focusing generates oscillating compressive strain in the
focal region; the oscillation of the optical scatterers is in the MHz frequency range.
Second, high-intensity focused ultrasound can also generate low-frequency
(of kHz range) strain in elastic media, which in turn generates shear wave
propagating in the medium. Both modifications of scatterers can cause decorrelation
of the optical speckle patterns. However, when the speckle patterns are recorded over exposure
time windows after the ultrasound focusing, the effect of the high-frequency compressive motion is negligible
 and the decorrelation in speckle patterns is dominated by the second mechanism. 
 Moreover, if  the exposure times are sufficiently short
with respect to the shear motion, one may consider that, for each acquisition, the speckle pattern is created 
by a frozen scattering medium in which the optical scatterers are displaced by the displacement field 
of the shear wave corresponding to the central time of the time window.
 Based on this
  idea, transient opto-elastography experiments have been carried out by Bossy {\it et al.}\cite{Bossy07, Bossy09}, 
  where they found qualitative relations between the
  decorrelation of speckle patterns and optical absorption as well as mechanical
  properties (such as Young's modulus) of soft biological tissues.

The main objective of this paper is to provide an analytical model that relates the decorrelation
of the transmitted speckle patterns to the displacements of the optical scatterers. Such a model hence helps us to
 understand the aforementioned "acousto-elasto-optic" phenomenon. To this end, we start with
the Helmholtz equation with the  index of refraction having a highly oscillatory random part.
 We consider the regime where the correlation length of the random medium is of the same order as the wavelength
 and both are smaller than the typical propagation distance. We denote by $\eps$ the ratio  between
 the correlation length of the medium and the propagation distance and assume therefore that $\eps \ll 1$.
 We also assume that   the relative amplitude of the random fluctuations of the index of refraction is weak, of order $\sqrt{\eps}$.
 This is known to be a scaling regime where the random medium interacts 
 with the propagating high-frequency waves\cite{Ryzhik96}. We consider
 two random media, with the second medium obtained by shifting  
 the first medium in some local region.

The correlation of wave fields is well described by the Wigner distributions\cite{LP93}. Following the techniques of Ref.~\onlinecite{Ryzhik96},
we formally derive the radiative transfer equations (RTEs) for the cross-correlations of the two wave fields acquired with
 two random media in the limit $\eps$ goes to zero. Radiative transfer limits for waves in two different random media have already been considered, e.g. in Refs.~\onlinecite{BalRam04,BalRyz05}, but the case of two media related by a local shift considered in this paper is new. 
The salient effects of this local shift of random media are:
it introduces a phase modulation to the scattering cross section of the RTE in the region
affected by the shift; further, when the amplitude of the shift is large, say much larger than the wavelength, by non-stationary phase the RTE for the cross-correlation function is intrinsically absorbing in this region. Next, following the techniques of Refs.~\onlinecite{LK74, BLP79,DLv6}, we  derive the diffusion approximation of the RTE in the regime
 when the mean free path is small; this simplification is useful for numerical simulations. In the special case of large shift, 
 the cross correlation vanishes in the region affected by the shift. We shall see that this accounts for the loss of correlation of the speckle patterns.

The  paper is organized as follows. In the next section, we present the model of two random media considered in this paper; they are related by a local shift. In Section \ref{sec:RTElimit}, we derive the RTEs for the Wigner distributions of the wave fields in the aforementioned two random media, in the limit when $\eps$ goes to zero; we also present the formula for speckle pattern correlation in terms of the solutions of these RTEs. In Section \ref{sec:DFlimit}, we derive the diffusion approximation of the RTEs in the limit when the mean free path goes to zero, and again we derive the formula for the speckle pattern correlation in the diffusion regime.
In the special case of large shift, using the derived diffusion approximation and correlation formula, we run numerical simulations which confirm that the diffusion equation model is able to capture the loss of correlation of the speckle patterns.

\section{Wave equation and heterogeneous media}

In the microscopic scale, 
light propagation is described by the Maxwell equations.
In a $d$-dimensional medium ($d\geq 2$),
when scalar approximation is valid, these equations reduce to the following
Helmholtz equation for the electric field $u$,
\begin{equation}
\Delta u(\bx) + k_0^2 n^2 (\bx) u({\bx}) = 0, \quad \bx \in \R^d,
\label{eq:helm}
\end{equation}
where $k_0$ is the wave number of the light in vacuum and the spatially varying refractive index $n(\bx)$
 models the heterogeneous medium.

The following model for $n^2(\bx)$ is adopted:
\begin{equation}
n^2(\bx) : = n_0^2\left(1+ 2\sigma V(\frac{\bx}{l}) \right),
\label{eq:rind}
\end{equation}
where $V(\bx)$ is a real-valued mean-zero random process.
Therefore, the square index of refraction of the heterogeneous medium has mean $n^2_0$, which is assumed to
be a constant, and its relative fluctuations are captured by $2\sigma V( {\bx}/{l})$
where the numbers $\sigma$ and $l$ model the strength and the correlation
length of the fluctuations respectively. 

The random process $V(\bx)$ is assumed to be stationary, {\it i.e.}, statistically homogeneous.
The two-point correlation function of this process is
\begin{equation}
R(\bx) = \E [V(\by) V(\by + \bx)] = \E [V({\bf 0})V(\bx)],
\label{eq:Rdef}
\end{equation}
where $\E$ stands for the expectation with respect to the distribution of the random medium.
Throughout this paper, we adopt the convention that the Fourier transform of a function
 $f$ on $\R^d$ is defined by
\begin{equation}
\hat{f}(\bxi) := \frac{1}{(2\pi)^d} \int e^{i\bk \cdot \bx} f(\bx) d\bx.
\label{eq:fourier}
\end{equation}
Using this notation and the stationarity of $V$, one easily verifies that
\begin{equation}
\E \big[ \hat{V} (\bp) \hat{V}(\bq)\big]  = \hat{R}(\bp) \delta(\bp + \bq),
\label{eq:FR}
\end{equation}
where $\delta$ is the Dirac distribution. 

From now on the product $k_0^2 n_0^2$ is denoted by $k^2$.
Let $L$ denote the typical propagation distance. In the high-frequency regime, the ratio between the
wavelength and the propagation distance, denoted by $\eps := k^{-1}/L$, is much smaller than one.
Let $\kappa:= l/k^{-1}$ be the ratio of the correlation length of the random medium to the
wavelength. We consider the so-called weak coupling high-frequency regime which corresponds
to $\kappa = 1$ and $\sigma = \sqrt{\eps}$. This is known to be a situation where the
heterogeneous medium interacts with the high-frequency waves\cite{Ryzhik96}.
Denoting by $u^\eps(\bx)$ the scaled function $u(L\bx)$, the Helmholtz equation becomes
\begin{equation}
\frac{\eps^2}{2} \Delta u^\eps (\bx) + \frac{1}{2} u^\eps ({\bx}) + \sqrt{\eps} V^\eps (\bx) u^\eps (\bx) = 0 ,
\label{eq:rhelm}
\end{equation}
with $V^\eps(\bx) = V(\bx/\eps)$.

{\bfseries Two random media.} As mentioned in the Introduction, we are interested in
the correlation of two wave fields in two random media. With the application to opto-elastography in mind, we view these two media as configurations of scatterers
at two instants of the shear motion. Let $u^\eps_1$ be the first wave field which
solves \eqref{eq:rhelm} with $V_1^\eps(\bx) = V( {\bx}/{\eps})$. Let $u^\eps_2$ be the second
wave field which solves \eqref{eq:rhelm} with $V_2^\eps(\bx)$ given by
\begin{equation}
V_2^\eps(\bx) = V_1^\eps\big( \bx +\eps \bphi(\bx) \big)  =  V\big(\frac{\bx}{\eps} + \bphi (\bx)\big), \quad \quad V_1^\eps(\bx) = V\big(\frac{\bx}{\eps}\big).
\label{eq:shift}
\end{equation}
Here $\bphi$ is a continuous and compactly supported vector field. $V_2^\eps$ is obtained from
$V_1^\eps$ by the application of the diffeomorphism $\bx \to \bx +\eps \bphi(\bx)$.
Hence $\bphi$ can be
thought as the displacement field that shifts the configurations of optical scatterers.
Note that the amplitude of the shift is of order $\eps$, which is of the order of the optical wavelength.

{\bfseries Correlation of waves.} It is well known that the correlation of two scalar wave fields $u$
and $v$ is well described by the Wigner distribution $W[u,v]$, which is defined by
\begin{equation}
W[u,v](\bx, \bk) : = \frac{1}{(2\pi)^d} \int e^{i\bk \cdot \by}
u (\bx - \frac{\eps \by}{2}) \overline{v} (\bx + \frac{\eps
\by}{2}) d\by,
\label{eq:Wigdef}
\end{equation}
where the bar denotes the complex conjugate. According to \eqref{eq:fourier},
the Wigner distribution may be thought as the Fourier transform of the two-point correlation function
of the fields. In this paper we are interested in
\begin{equation}
W^\eps_{jl}(\bx, \bk) = W[u^\eps_j, u^\eps_l] (\bx, \bk),
\label{eq:wigner}
\end{equation}
where the wave fields $u^\eps_j$, $j = 1, 2$, are described earlier and
they correspond to two random media $V_j^\eps$ related by \eqref{eq:shift}. We note that
\begin{equation}
\int W^\eps_{jl}(\bx,\bk) d\bk = u^\eps_j(\bx) \overline{u^\eps_l}(\bx).
\label{eq:Wkint}
\end{equation}
In particular, $\int W^\eps_{11} (\bx,\bk) d\bk$ is the energy density $|u^\eps_1(\bx)|^2$
and $W^\eps_{11} (\bx,\bk)$ itself corresponds to frequency-specified energy density
of the first wave field. The functions $W^\eps_{22}$ and $W^\eps_{12}$ can be interpreted similarly.

In the next section, we derive the radiative transfer equation (RTE) for $W^\eps_{jl}$
from the Helmholtz equation
 in the limit $\eps \to 0$, and we relate the correlation of speckle patterns to the Wigner distributions.

\section{Radiative transport equation for the Wigner distributions}
\label{sec:RTElimit}

The goal of this section is to derive the RTE for the correlation of waves.
Recall that the two wave fields $u^\eps_j$, $j=1,2$, solve
\begin{equation}
\frac{\eps^2}{2} \Delta u^\eps_j (\bx) + \frac{1}{2} u^\eps_j ({\bx}) + \sqrt{\eps} V_j^\eps (\bx)u^\eps_j (\bx) = 0,
\label{eq:uepsjl}
\end{equation}
where $V_2^\eps$ and $V_1^\eps$ are of the form \eqref{eq:shift}. Recall that $W^\eps_{jl}$ is defined in \eqref{eq:wigner}. For the moment, we assume that $u^\eps_j$ has prescribed plane wave behavior at infinity to simplify the calculation and derive RTEs for the limits of $W^\eps_{jl}$'s. For bounded domains, these RTEs are valid in the interior of the domain, and appropriate boundary conditions should be imposed (that  will be discussed later).

Using \eqref{eq:wigner} and \eqref{eq:uepsjl}, integration by parts and the prescribed plane wave behaviors of $u^\eps_j$'s, we find
(see Appendix \ref{app:A} for the derivation):
\begin{eqnarray}
\label{wigner11eps}
i\bk \cdot \nabla W^\eps_{11}(\bx,\bk) &=& \frac{1}{\sqrt{\eps}} \int \frac{e^{i\bk \cdot \by}}{(2\pi)^d} \left[V(\frac{\bx}{\eps} + \frac{\by}{2}) - V(\frac{\bx}{\eps} - \frac{\by}{2})\right] u^\eps_1(\bx-\frac{\eps \by}{2}) \overline{u^\eps_1}(\bx+\frac{\eps \by}{2}) d\by,\\
\nonumber
i\bk \cdot \nabla W^\eps_{12}(\bx,\bk) &=& \frac{1}{\sqrt{\eps}} \int \frac{e^{i\bk \cdot \by}}{(2\pi)^d} \left[V(\frac{\bx}{\eps} + \frac{\by}{2} + \bphi (\bx + \frac{\eps\by}{2})) - V(\frac{\bx}{\eps} - \frac{\by}{2})\right] \\
\label{wigner12eps}
&& \hspace*{0.9in} \times
u^\eps_1(\bx-\frac{\eps \by}{2}) \overline{u^\eps_2} (\bx+\frac{\eps \by}{2})d\by,\\
\nonumber
i\bk \cdot \nabla W^\eps_{22}(\bx,\bk) &=& \frac{1}{\sqrt{\eps}} \int \frac{e^{i\bk \cdot \by}}{(2\pi)^d} \left[V(\frac{\bx}{\eps} + \frac{\by}{2} + \bphi (\bx + \frac{\eps\by}{2})) - V(\frac{\bx}{\eps} - \frac{\by}{2} + \bphi (\bx - \frac{\eps\by}{2}))\right] \\
\label{wigner22eps}
&& \hspace*{0.9in}\times u^\eps_2
(\bx-\frac{\eps \by}{2})\overline{u^\eps_2}(\bx+\frac{\eps \by}{2}) d\by.
\end{eqnarray}

Formal derivation of RTE for $W^\eps_{11}$ is classic\cite{Ryzhik96}. 
We observe, however, that the equation for
$W^\eps_{12}$ and $W^\eps_{22}$ are not standard because of the
shift $\bphi$. In fact, correlation of wave fields in different
random media has been considered in Ref.~\onlinecite{BalRam04}, but they do
not cover the case when one medium is a local shift of the other
as described in \eqref{eq:shift}.

Now we adapt the formal derivations in Refs.~\onlinecite{Ryzhik96,BalRam04} to
derive RTE for the cross-correlations $W^\eps_{12}$ and
$W^\eps_{22}$. In the process of the derivation and as in the
references, various assumptions will be made whose rigorous
justifications are known to be difficult. To form a closed
equation for $W^\eps_{12}$, we rewrite the equation it satisfies
as follows using the Fourier transform of $V$,
\begin{eqnarray*}
i\bk \cdot \nabla W^\eps_{12}(\bx,\bk) = \frac{1}{\sqrt{\eps}} \iint \frac{e^{i\bk \cdot \by} d\by d\bp}{(2\pi)^d} \hat{V}(\bp) \left[e^{-i\bp \cdot \left(\frac{\bx}{\eps} + \frac{\by}{2} + \bphi (\bx + \frac{\eps \by}{2})\right)} - e^{-i\bp \cdot \left(\frac{\bx}{\eps} - \frac{\by}{2}\right)}\right] \\
\times
u^\eps_1(\bx-\frac{\eps \by}{2}) \overline{u^\eps_2}(\bx+\frac{\eps \by}{2}).
\end{eqnarray*}
We  replace the function $\bphi (\bx + \eps \by/2)$ in the
phase function by $\bphi(\bx)$ and neglect the error term of order $\eps$. It
follows that
\begin{eqnarray}
\nonumber
i\bk \cdot \nabla W^\eps_{12}(\bx,\bk) &\approx& \frac{1}{\sqrt{\eps}} \iint \frac{e^{i\bk \cdot \by} d\by d\bp}{(2\pi)^d} \hat{V}(\bp) \left[e^{-i\bp \cdot \left(\frac{\bx}{\eps} + \frac{\by}{2} + \bphi (\bx)\right)} - e^{-i\bp \cdot \left(\frac{\bx}{\eps} - \frac{\by}{2}\right)}\right] \\
\nonumber
&& \hspace*{1.7in} \times
u^\eps_1(\bx-\frac{\eps \by}{2}) \overline{u^\eps_2}(\bx+\frac{\eps \by}{2})\\
&=& \frac{1}{\sqrt{\eps}} \int  e^{-i\bp \cdot \frac{\bx}{\eps}} \hat{V}(\bp)\left[e^{-i\bp \cdot \bphi (\bx)} W^\eps_{12}(\bx,\bk-\frac{\bp}{2}) -  W^\eps_{12}(\bx,\bk+\frac{\bp}{2}) \right] d\bp,
\label{eq:W12eps:a}
\end{eqnarray}
where in the second equality we have used the definition \eqref{eq:wigner}. By the same argument we
also have
\begin{equation}
i\bk \cdot \nabla W^\eps_{22}(\bx,\bk) \approx \frac{1}{\sqrt{\eps}} \int e^{-i\bp \cdot \left(\frac{\bx}{\eps}
+ \bphi(\bx)\right)} \hat{V}(\bp) \left[W^\eps_{22}(\bx,\bk-\frac{\bp}{2}) - W^\eps_{22}(\bx,\bk+\frac{\bp}{2}) \right]d\bp.
\label{eq:W22eps:a}
\end{equation}

\subsection{Radiative transfer limit by multiscale expansion}
\label{sec:rte}%
In this subsection we consider the limit of $W^\eps_{12}$ and $W^\eps_{22}$ as $\eps$ goes to zero using a
formal multiscale expansion argument. The equations (\ref{eq:W12eps:a}) and (\ref{eq:W22eps:a}) can be written as
\begin{equation}
\bk \cdot \nabla W^\eps_{jl} (\bx, \bk) + \frac{1}{\sqrt{\eps}} \mathcal{P}_{jl} W^\eps_{jl}(\bx,\bk) = 0.
\label{eq:Eqforexp}
\end{equation}
Here $jl$ takes the value $12$ or $22$. The operators $\mathcal{P}_{jl}$ are defined by
\begin{equation*}
\begin{aligned}
\mathcal{P}_{12} W^\eps_{12} &= i\int  e^{-i\bp \cdot \frac{\bx}{\eps}} \hat{V}(\bp) \left[ e^{-i\bp \cdot \bphi(\bx)} W^\eps_{12}(\bx,\bk-\frac{\bp}{2}) -  W^\eps_{12}(\bx,\bk+\frac{\bp}{2}) \right] d\bp,\\
\mathcal{P}_{22} W^\eps_{22} &= i \int  e^{-i\bp \cdot \left(\frac{\bx}{\eps} + \bphi(\bx)\right)} \hat{V}(\bp)\left[ W^\eps_{22}(\bx,\bk-\frac{\bp}{2}) - W^\eps_{22}(\bx,\bk+\frac{\bp}{2}) \right] d\bp.
\end{aligned}
\end{equation*}

Following the method of multiscale expansion, we define the fast variable $\bxi = \bx/\eps$,
and assume that the following expansion is valid:
\begin{equation}
W^\eps_{jl}(\bx,\bk) = W^{(0)}_{jl}(\bx,\bxi,\bk) + \sqrt{\eps} W^{(1)}_{jl}(\bx,\bxi,\bk) + \eps W^{(2)}_{jl}(\bx,\bxi,\bk) + \cdots
\label{eq:mansatz}
\end{equation}
As $\eps$ goes to zero, the leading-order term $W^{(0)}_{jl}$ dominates, and the goal is to derive a
RTE for this term. To this end, we  substitute the multiscale ansatz \eqref{eq:mansatz} and the relation
\begin{equation*}
\nabla = \nabla_{\bx} + \frac{1}{\eps} \nabla_{\bxi}
\end{equation*}
into \eqref{eq:Eqforexp}. Equating terms that are of equal order in $\eps$, we find:
\begin{eqnarray}
&\mathcal{O}(\eps^{-1}): \quad & \bk \cdot \nabla_{\bxi} W^{(0)}_{jl} = 0,
\label{eq:rteexp0} \\
&\mathcal{O}(\eps^{-1/2}): \quad & \bk \cdot \nabla_{\bxi} W^{(1)}_{jl} + \mathcal{P}_{jl} W^{(0)}_{jl} = 0,
\label{eq:rteexp1}\\
&\mathcal{O}(1): \quad & \bk\cdot \nabla_{\bx} W^{(0)}_{jl} + \bk \cdot \nabla_{\bxi} W^{(2)}_{jl} + \mathcal{P}_{jl} W^{(1)}_{jl} = 0.
\label{eq:rteexp2}
\end{eqnarray}
The strategy of the derivation is as follows: \\
- The first equation imposes that $W^{(0)}_{jl}$ is
 independent of the fast variable $\bxi$. \\
 - The second equation can be explicitly inverted and
 provides a representation of $W^{(1)}_{jl}$ in terms of $W^{(0)}_{jl}$. \\
 - By taking the statistical
expectation of the third equation, we get a compatibility equation that gives the equation satisfied by $W^{(0)}_{jl}$.
Indeed, we observe that
\begin{equation*}
\E \big[  \nabla_{\bxi} W^{(2)}_{jl} \big]  = 0,
\end{equation*}
because the statistics of $V$ (and therefore of $W^{(2)}_{jl}$) is homogeneous in the
fast variable $\bxi$. We
also note  that it is natural to assume that $W^{(0)}_{jl}$ is deterministic. Then the expectation of the third equation reduces to
\begin{equation}
\bk\cdot \nabla_{\bx}  W^{(0)}_{jl}  + \E \big[ \mathcal{P}_{jl} W^{(1)}_{jl} \big]  = 0.
\end{equation}
Therefore, it suffices to invert \eqref{eq:rteexp1} and to evaluate the expectation above.

{\bf Limiting equation for $W^\eps_{12}$}.  Upon adding an absorption term $\theta W^{(1)}_{12}$
for regularization (that will be set to zero later), we rewrite \eqref{eq:rteexp1} as
\begin{equation}
\label{eq:rteexp1bis}
\bk \cdot \nabla_\bxi W^{(1)}_{12} + \theta W^{(1)}_{12} +  i \int e^{-i\bp \cdot \bxi} \hat{V}(\bp) \left[ e^{-i\bp \cdot \bphi (\bx)} W^{(0)}_{12}(\bk - \frac{\bp}{2}) - W^{(0)}_{12}(\bk + \frac{\bp}{2}) \right] d\bp = 0.
\end{equation}
Taking Fourier transform in the fast variable $\bxi$:
$$
\hat{W}^{(1)}_{12} (\bx, \bq, \bk) = \frac{1}{(2\pi)^d} \int  {W}^{(1)}_{12} (\bx, \bxi, \bk) e^{ i \bq \cdot \bxi} d\bxi,
$$ 
and making use of
$$
\frac{1}{(2\pi)^d} \int \nabla_\bxi {W}^{(1)}_{12} (\bx, \bxi, \bk) e^{ i \bq \cdot \bxi} d\bxi = -i \bq \hat{W}^{(1)}_{12} (\bx, \bq, \bk),
$$
we obtain from (\ref{eq:rteexp1bis}) that
\begin{equation}
\hat{W}^{(1)}_{12} (\bx, \bq, \bk) = \frac{\hat{V}(\bq)\left[ e^{-i\bq \cdot \bphi (\bx)} W^{(0)}_{12}(\bk - \frac{\bq}{2}) - W^{(0)}_{12}(\bk + \frac{\bq}{2}) \right]}{\bk \cdot \bq + i\theta}.
\label{eq:FTW121}
\end{equation}
To estimate the expectation of $\mathcal{P}_{12} W^{(1)}_{12}$, we rewrite this integral in terms of the Fourier
transform of $W^{(1)}_{12}$:
\begin{equation*}
\mathcal{P}_{12} W^{(1)}_{12}(\bx,\bxi,\bk) =  i\iint  e^{-i(\bp + \bq) \cdot \bxi} \hat{V}(\bp) \left[ e^{-i\bp \cdot \bphi(\bx)} \hat{W}^{(1)}_{12}(\bx,\bq,\bk-\frac{\bp}{2}) -  \hat{W}^{(1)}_{12}(\bx,\bq,\bk+\frac{\bp}{2}) \right] d\bq d\bp.
\end{equation*}
Using \eqref{eq:FTW121}, we can write this integral in terms of $W^{(0)}_{12}$ as follows:
\begin{equation*}
\begin{aligned}
\mathcal{P}_{12} W^{(1)}_{12}(\bx,\bxi,\bk) = & i\iint e^{-i(\bp + \bq)\cdot \bxi} \hat{V}(\bp)\hat{V}(\bq) \\
&\times \left\{  \frac{e^{-i\bp \cdot \bphi (\bx)} 
 \left[ e^{-i\bq \cdot \bphi (\bx)} W^{(0)}_{12}(\bk - \frac{\bp}{2} - \frac{\bq}{2}) - W^{(0)}_{12}(\bk - \frac{\bp}{2} + \frac{\bq}{2}) \right]}{(\bk - \frac{\bp}{2}) \cdot \bq + i\theta} \right.\\
&\quad  \left. - \frac{e^{-i\bq \cdot \bphi (\bx)} W^{(0)}_{12}(\bk + \frac{\bp}{2} - \frac{\bq}{2}) - W^{(0)}_{12}(\bk + \frac{\bp}{2} + \frac{\bq}{2})}{(\bk+\frac{\bp}{2}) \cdot \bq + i\theta}\right\} d\bp d\bq.
\end{aligned}
\end{equation*}
Taking expectation, using the definition \eqref{eq:FR} and the fact that $\hat{R}(\bp) = \hat{R}(-\bp)$,
we obtain that the expectation of $\mathcal{P}_{12} W^{(1)}_{12}$ can be evaluated as
\begin{equation*}
\begin{aligned}
& \E \big[ \mathcal{P}_{12} W^{(1)}_{12} \big] \\
& = i\int  \hat{R}(\bp) \left\{  \frac{W^{(0)}_{12}(\bk) - e^{-i\bp \cdot \bphi (\bx)} W^{(0)}_{12}(\bk - \bp)}{-(\bk - \frac{\bp}{2}) \cdot \bp + i\theta} - \frac{e^{i\bp \cdot \bphi (\bx)} W^{(0)}_{12}(\bk + \bp) - W^{(0)}_{12}(\bk)}{-(\bk+\frac{\bp}{2}) \cdot \bp + i\theta}\right\} d\bp\\
&= i \int  \hat{R}(\bk - \bp) \frac{W^{(0)}_{12}(\bk) - e^{-i(\bk-\bp) \cdot \bphi (\bx)} W^{(0)}_{12}(\bp)}{-\frac{1}{2}(|\bk|^2 - |\bp|^2) + i\theta} + \hat{R}(\bp - \bk) \frac{e^{i(\bp-\bk) \cdot \bphi  (\bx)} W^{(0)}_{12}(\bp) - W^{(0)}_{12}(\bk)}{- \frac{1}{2}(|\bk|^2 - |\bp|^2) -i\theta}d\bp\\
&= i \int  \hat{R}(\bp - \bk) [W^{(0)}_{12}(\bk) - W^{(0)}_{12}(\bp) e^{i(\bp - \bk)\cdot \bphi (\bx)}] \frac{-2i\theta}{\frac{1}{4}(|\bk|^2 - |\bp|^2)^2 + \theta^2} d\bp\\
&  \xrightarrow{\theta \to 0}  4\pi \int  \hat{R}(\bp - \bk) [W^{(0)}_{12}(\bk) - W^{(0)}_{12}(\bp) e^{i(\bp - \bk)\cdot \bphi (\bx)}] \delta(|\bk|^2 - |\bp|^2) d\bp.
\end{aligned}
\end{equation*}
In the first equality, we have used the change of variables $\bk - \bp \mapsto \bp$ and $\bk + \bp \mapsto \bp$ for
the two integrands, respectively. In the last equality, we have used the fact that
\begin{equation*}
\frac{\theta}{x^2 + \theta^2} \xrightarrow{\theta \to 0} \pi \delta(x)
\end{equation*}
as a distribution of the one-dimensional variable $x$.
Finally, we obtain the following RTE for $W^{(0)}_{12}$:
\begin{equation}
\label{eq:rte12}
\bk \cdot \nabla W^{(0)}_{12} (\bx,\bk) + 4\pi \int \hat{R}(\bp - \bk) \big[W^{(0)}_{12}(\bk) - W^{(0)}_{12}(\bp) e^{i(\bp - \bk)\cdot \bphi (\bx)} \big]
 \delta\big(|\bk|^2 - |\bp|^2 \big) d\bp = 0.
\end{equation}

{\bf Limiting equation for $W^\eps_{22}$}. The above procedure
can be applied to the equation satisfied by $W^\eps_{22}$ as well. Upon
adding a regularizing term and using Fourier transform, we can solve
equation \eqref{eq:rteexp1}:
\begin{equation}
\hat{W}^{(1)}_{22} (\bx, \bq, \bk) = \frac{\hat{V}(\bq)e^{-i\bq \cdot \bphi (\bx)} \left[ W^{(0)}_{22}(\bk - \frac{\bq}{2}) - W^{(0)}_{22}(\bk + \frac{\bq}{2}) \right]}{\bk \cdot \bq + i\theta}.
\label{eq:FTW221}
\end{equation}
Using this solution and the Fourier transform representation, we find that $\mathcal{P}_{22} W^{(1)}_{22}$ has the form
\begin{equation*}
\begin{aligned}
\mathcal{P}_{22} W^{(1)}_{22}(\bx,\bxi,\bk) =  i \iint e^{-i(\bp + \bq)\cdot (\bxi + \bphi(\bx))} \hat{V}(\bp)\hat{V}(\bq)  \left\{  \frac{W^{(0)}_{22}(\bk - \frac{\bp}{2} - \frac{\bq}{2}) - W^{(0)}_{22}(\bk - \frac{\bp}{2} + \frac{\bq}{2})}{(\bk - \frac{\bp}{2}) \cdot \bq + i\theta} \right.\\
 \left. - \frac{W^{(0)}_{22}(\bk + \frac{\bp}{2} - \frac{\bq}{2}) - W^{(0)}_{22}(\bk + \frac{\bp}{2} + \frac{\bq}{2})}{(\bk+\frac{\bp}{2}) \cdot \bq + i\theta}\right\} d\bp d\bq.
\end{aligned}
\end{equation*}
This expression is much simpler compared with that of
$\mathcal{P}_{12} W^{(1)}_{12}$ because the phase modification due
to $\bphi$ is uniform for the $W^{(0)}_{22}$ above. Taking
expectation and using \eqref{eq:FR} we find that this modification
has no effect on the expectation. In fact, we have
\begin{equation*}
\E \big[ \mathcal{P}_{22} W^{(1)}_{22}(\bx,\bxi,\bk)\big]  \rightarrow  4\pi \int  \hat{R}(\bp - \bk) \big[W^{(0)}_{22}(\bk) - W^{(0)}_{22}(\bp) \big]
 \delta \big(|\bk|^2 - |\bp|^2\big) d\bp.
\end{equation*}

Finally, we obtain the RTE for $W^{(0)}_{22}$:
\begin{equation}
\label{eq:rte22}
\bk \cdot \nabla W^{(0)}_{22} (\bx,\bk) + 4\pi \int \hat{R}(\bp - \bk) \big[W^{(0)}_{22}(\bk) - W^{(0)}_{22}(\bp) \big] \delta \big(|\bk|^2 - |\bp|^2\big) d\bp = 0.
\end{equation}
We note that this is the same as the classic RTE limit for $W^{(0)}_{11}$.

{\bf Summary of the results.} Now we summarize the above results
and discuss the boundary conditions when the problem is posed on a
bounded domain as often encountered in practice. 


As can be seen from the presence of the Dirac term in (\ref{eq:rte12}) and (\ref{eq:rte22}) there is no coupling between 
Wigner distributions with different values of $|\bk|$. From now on
we consider the monokinetic case
when $|\bk|>0$ is fixed, so the Wigner distribution is specified by the
spatial variable $\bx \in X$
 and the direction $\hat{\bk} \in S^{d-1}$:
\begin{equation}
W_{jl}(\bx,\hat{\bk})= W_{jl}^{(0)}(\bx,|\bk|\hat{\bk}) ,
\end{equation} 
where $X$ is the physical domain and $S^{d-1}$ is the unit sphere in $\R^d$.
 The RTEs are therefore posed on the phase space $(\bx,\hat{\bk}) \in X \times S^{d-1}$, with $|\bk|$
 as a fixed parameter.
 With these notations, the equations \eqref{eq:rte12} and \eqref{eq:rte22} become
\begin{eqnarray}
\hat{\bk} \cdot \nabla W_{12}(\bx,\hat{\bk}) + \Sigma ( \hat{\bk};|\bk|)  W_{12} (\bx,\hat{\bk}) &=& \int_{S^{d-1}} \sigma (\hat{\bp}, \hat{\bk};|\bk|) e^{i|\bk| (\hat{\bp}-\hat{\bk})\cdot \bphi(\bx)} W_{12}(\bx, \hat{\bp}) d\hat{\bp}, \label{eq:rte12it}\\
\hat{\bk} \cdot \nabla W_{jj} (\bx,\hat{\bk}) + \Sigma ( \hat{\bk};|\bk|)  W_{jj} (\bx,\hat{\bk}) &=& \int_{S^{d-1}} \sigma (\hat{\bp}, \hat{\bk};|\bk|)  W_{jj}(\bx,  \hat{\bp}) d\hat{\bp}, \quad j=1,2,\label{eq:rte22it}
\end{eqnarray}
for $(\bx,\hat{\bk}) \in X \times S^{d-1}$.
In (\ref{eq:rte12it}-\ref{eq:rte22it}) the differential scattering cross section $\sigma$ is defined by 
\begin{equation}
\sigma(\hat{\bp},\hat{\bk}; |\bk|) = 2\pi|\bk|^{d-3} \hat{R}\big( (\hat{\bp} - \hat{\bk}) |\bk| \big), \quad \hat{\bp}, \hat{\bk} \in S^{d-1} ,
\label{eq:scatKer}
\end{equation}
and the total scattering cross section $\Sigma$ is defined by:
\begin{equation}
\Sigma ( \hat{\bk};|\bk|) = \int_{S^{d-1}} \sigma (\hat{\bp}, \hat{\bk}; |\bk|) d\hat{\bp},\quad  \hat{\bk} \in S^{d-1}  .
\label{eq:tabs}
\end{equation}
The total scattering cross section is such that
$$
 \frac{4\pi}{|\bk|} \int \delta(|\bk|^2 - |\bp|^2) \hat{R}(\bp -\bk) d\bp=\Sigma ( \hat{\bk};|\bk|) ,
$$
because
\begin{equation*}
\delta(|\bk|^2 - |\bp|^2) = \frac{1}{2|\bk|} \delta(|\bk|-|\bp|).
\end{equation*}

On a bounded domain $X \subset \R^d$, we need to equip the RTEs (\ref{eq:rte12it}-\ref{eq:rte22it})
with proper boundary conditions. 
 We define the incoming boundary $\Gamma_-$ and the outgoing boundary $\Gamma_+$ as
\begin{equation}
\Gamma_{\pm} := \{(\bx, \hat{\bk}) \in \partial X \times S^{d-1}~|~  \pm \hat{\bk}  \cdot \bnu(\bx) > 0 \},
\label{eq:Gammapm}
\end{equation}
where $\bnu(\bx)$ is the outer-pointing normal at the point $\bx$ on the boundary $\partial X$.
When a bounded domain is considered, the RTEs \eqref{eq:rte12it} and \eqref{eq:rte22it} for the
Wigner distributions $W_{jl}$'s should be understood as for $(\bx, \hat{\bk}) \in X \times S^{d-1}$ with the 
boundary condition
\begin{equation}
W_{jl}(\bx, \hat{\bk}) = p(\bx, \hat{\bk}), \quad (\bx, \hat{\bk}) \in \Gamma_-,
\label{eq:RTEBC}
\end{equation}
where $p$ models the light intensity incoming at point $\bx \in \partial X$ in the direction $\hat{\bk}$.
In the case of an incident laser beam the support of $p$ is spatially limited to $\bx \in \partial X_{\rm i} $
where $\partial X_{\rm i}$ is the part of $\partial X$ where the incident laser beam is applied.

{\bfseries The case of large shifts.} 
When the shift $\bphi$ is much larger than the wavelength, i.e.,
\begin{equation}
|\bk| |\bphi| \gg 1,
\end{equation}
the RTE (\ref{eq:rte12it}) for $W_{12}$ can be simplified. Indeed, since the integral
is taken over the sphere which has non-vanishing Gaussian
curvature, we can apply  Theorem 1.2.1 in Ref.~\onlinecite{Sogge93}, which may be
viewed as an analog of the Riemann-Lebesgue lemma, and conclude
that the integral on the right-hand side of \eqref{eq:rte12it} is of
order $(|\bk||\bphi|)^{-(d-1)/2}$ and hence approaches zero.
Consequently, \eqref{eq:rte12it} should be modified as follows
\begin{equation}
\begin{aligned}
\hat{\bk} \cdot \nabla W_{12} (\bx,\hat{\bk}) + \Sigma( \hat{\bk};|\bk|)  W_{12}(\bx,\hat{\bk})  &= 0, \quad & (\bx,\hat{\bk}) \in X_{\mathrm{s}} \times S^{d-1}, \\
\hat{\bk} \cdot \nabla W_{12} (\bx,\hat{\bk}) + \Sigma( \hat{\bk};|\bk|)  W_{12} (\bx,\hat{\bk}) &= \int_{S^{d-1}} \sigma (\hat{\bp}, \hat{\bk};|\bk|)  W_{12}(\bx, \hat{\bp}) d\hat{\bp}, \quad & (\bx,\hat{\bk}) \in X^c_{\mathrm{s}} \times S^{d-1}.\label{eq:rte12ls}
\end{aligned}
\end{equation}
Here $X_{\mathrm{s}}$ is the support of $\bphi$, {\it i.e.}, the region in which the scatterers are shifted,
and $X^c_{\mathrm{s}}$ is the complement of $X_{\mathrm{s}}$ in $X$.

\subsection{Speckle pattern correlations in the RTE regime}

As we have seen, the Wigner distribution $W[u,u]$ of a wave field $u$ has the interpretation of direction-resolved 
energy density. Hence RTE for $W[u,u]$ is a very good model for light propagation.
In this subsection, we relate the correlation of optical speckle patterns to  integrals of
Wigner distributions.

In the opto-elastography experiment, emitted light intensity is measured at a part $\partial X_{\rm m}$
of the domain boundary $\partial X$  and the data are $\{|u^{\mathrm{e}}_{j} (\bx)|^2 ~|~ \bx \in \partial X_{\rm m}\}$, $j=1,2$. The correlation of
two speckle patterns $u^{\mathrm{e}}_1$ and $u^{\mathrm{e}}_2$ is defined by
\begin{equation}
C_{12} = \frac{\langle (|u^{\mathrm{e}}_1|^2 - \langle |u^{\mathrm{e}}_1|^2 \rangle)(|u^{\mathrm{e}}_2|^2 - \langle |u^{\mathrm{e}}_2|^2 \rangle)\rangle}{\sqrt{\langle (|u^{\mathrm{e}}_1|^2 - \langle |u^{\mathrm{e}}_1|^2 \rangle)^2 \rangle} \sqrt{\langle (|u^{\mathrm{e}}_2|^2 - \langle |u^{\mathrm{e}}_2|^2 \rangle)^2 \rangle}},
\label{eq:C12def}
\end{equation}
where $\langle A \rangle$ denotes the spatial average over the boundary $\partial X_{\rm m}$, that is
\begin{equation}
\langle A \rangle = \frac{1}{|\partial X_{\rm m}|} \int_{\partial X_{\rm m}} A(\bx) d\bx.
\end{equation}
Note that in the above equation, the  notation $d\bx$ means the induced Lebesgue measure on the
 boundary $\partial X_{\rm m}$, and $|\partial X_{\rm m}|$ is the area of the boundary.

Assume that the complex amplitudes $(u^{\mathrm{e}}_1, u^{\mathrm{e}}_2)$, as a $\C^2$-valued random
process, satisfy the circular symmetric Gaussian distribution, and assume also that the spatial average
can be thought as ensemble averages (taking expectations), we have
\begin{equation}
\langle |u^{\mathrm{e}}_j|^2 |u^{\mathrm{e}}_l|^2 \rangle = \langle u^{\mathrm{e}}_j \overline{u^{\mathrm{e}}_l} \rangle \langle \overline{u^{\mathrm{e}}_j} u^{\mathrm{e}}_l\rangle + \langle |u^{\mathrm{e}}_j|^2\rangle \langle |u^{\mathrm{e}}_l|^2 \rangle, \quad j,l = 1,2.
\end{equation}
Using this equality, we find that the numerator of \eqref{eq:C12def} is
\begin{equation*}
\lvert \langle u^{\mathrm{e}}_1 \overline{u^{\mathrm{e}}_2} \rangle \rvert^2 = 
\frac{1}{|\partial X_{\rm m}|^2} \bigg| \int_{\partial X_{\rm m}}  u^{\mathrm{e}}_1(\bx) \overline{u^{\mathrm{e}}_2}(\bx) d\bx \bigg|^2  = \frac{1}{|\partial X_{\rm m}|^2} 
\bigg| \int_{\Gamma_{{\rm m},+}} W_{12}(\bx, \hat{\bk}) d\hat{\bk} d\bx \bigg|^2,
\end{equation*}
where 
$$
\Gamma_{{\rm m},+}  := \{(\bx, \hat{\bk}) \in \partial X \times S^{d-1} ~|~ \bx \in \partial X_{\rm m}, \, \hat{\bk}  \cdot \bnu(\bx) > 0 \}.
$$
In the second equality above, we  have used the fact that $u_1^{\mathrm{e}}\overline{u_2^{\mathrm{e}}}(\bx)$ is the integral over $\hat{\bk}$ of $W_{12}(\bx,\hat{\bk})$, as seen in \eqref{eq:Wkint}. Since the product $u^{\mathrm{e}}_1 \overline{u_2^{\mathrm{e}}}(\bx)$ only accounts for the outgoing light, we take the integral of $W_{12}$ over outgoing directions. 
 Similarly, the denominator in \eqref{eq:C12def} is given by
\begin{equation*}
\langle |u^{\mathrm{e}}_1|^2 \rangle  \langle |u^{\mathrm{e}}_2|^2 \rangle  =  \frac{1}{|\partial X_{\rm m}|^2} \int_{\Gamma_{{\rm m},+}}  W_{11}(\bx, \hat{\bk}) d\hat{\bk} d\bx \ \int_{\Gamma_{{\rm m},+}}  W_{22}(\bx, \hat{\bk}) d\hat{\bk} d\bx.
\end{equation*}
Combining these calculations above, we find that in the RTE regime, the correlation of two speckle patterns is:
\begin{equation}
C_{12} = \frac{\displaystyle \Big| \int_{\Gamma_{{\rm m},+}}  W_{12}(\bx, \hat{\bk}) d\hat{\bk} d\bx \Big| ^2}{\displaystyle
 \int_{\Gamma_{{\rm m},+}}  W_{11}(\bx, \hat{\bk}) d\hat{\bk} d\bx \ \int_{\Gamma_{{\rm m},+}}  W_{22}(\bx, \hat{\bk}) d\hat{\bk} d\bx}.
\label{eq:C12formRTE}
\end{equation}

We remark that the modeling of the outgoing light depends on the measurement set-up.
In the above model, the  light outgoing in all directions is completely captured
(hence the integral in $\hat{\bk}$ over the half-sphere $\bnu(\bx) \cdot \hat{\bk} >0$). 
If light is collected and imaged by a lens, then we should only take the integral
in $\hat{\bk}$ in a cone that corresponds to the aperture of the lens. Such a situation
does not affect qualitatively the results in the sequel.

\section{The diffusion limit for the radiative transfer equations}
\label{sec:DFlimit}

The RTEs (\ref{eq:rte12it}) and (\ref{eq:rte22it}) derived in the previous section are posed on the phase space $X \times S^{d-1}$,
which is of dimension $2d-1$. This may cause difficulties for instance for numerical simulations, especially when the total scattering cross section $\Sigma$ is large. In this section, we consider the diffusion approximations of the RTEs  (\ref{eq:rte12it}) and (\ref{eq:rte22it})
which are much easier to deal with for the purpose of simulations.

For the sake of simplicity, we consider the case when the differential scattering cross section $\sigma$, as defined in \eqref{eq:scatKer},
takes the form
\begin{equation}
\label{eq:sigmaiso}
\sigma(\hat{\bp},\hat{\bk}; |\bk|) = \sigma(\hat{\bp} \cdot \hat{\bk}; |\bk|).
\end{equation}
This happens in particular when the fluctuations of the index of refraction are statistically isotropic, {\it i.e.} when
$\hat{R}(\bp)$ depends only on $|\bp|$.
If (\ref{eq:sigmaiso}) holds, then  the total scattering cross section $\Sigma$ defined in \eqref{eq:tabs} depends only on the parameter $|\bk|$.
In the sequel of this paper, we will use the normalized differential scattering cross section
\begin{equation}
f(\hat{\bp}\cdot \hat{\bk};|\bk|) = \frac{1}{\Sigma(|\bk|)}\sigma(\hat{\bp} \cdot \hat{\bk}; |\bk|).
\end{equation}
 The diffusion approximation
is valid in the regime where the mean free path $\eta = 1/\Sigma(|\bk|)$ is much smaller than the size of the
domain. Here we assume $\eta \ll 1$ 
and the size of the domain is of order one (Alternatively, when $\eta$ is bounded but not necessarily small, diffusion limit is still valid when we consider the problem on a very large domain and rescale the spatial variable).

In the following two subsections, we adapt the classical derivation of diffusion limits of
 RTE\cite{LK74,BLP79,DLv6} to the case considered in this paper. The point here is that we have
 to manage the shift field $\bphi$. Though rigorous derivation of diffusion limit is possible,
 we do not pursue it here and adopt the formal multiscale expansion argument only.
 We consider three interesting situations. In the first one, the amplitude of the shift $\bphi$
  is very small so that $|\bk| |\bphi|$ is of order $\eta$. In the second one,
  the amplitude of  the shift $\bphi$ is large so that $|\bk| |\bphi|$ is of order one.
   In the third one, the amplitude of
  the shift $\bphi$ is very large so that $|\bk| |\bphi|$ is much larger than one. As seen in the previous section, this leads to the RTE \eqref{eq:rte12ls}.

\subsection{The case of small shifts}
\label{sec:ss}

In this subsection, the amplitude of the shift $\bphi$ is assumed to be small compared to the wavelength so that 
\begin{equation}
|\bk| \bphi(\bx)
= \eta \bpsi(\bx),
\end{equation}
for some function $\bpsi(\bx)$ of order one.
 The diffusion limit for $W_{11}$ and $W_{22}$ are classic (see Section XXI.5.4 in Ref.~\onlinecite{DLv6});
hence we concentrate on that for $W_{12}$ in (\ref{eq:rte12it}). Dividing on both sides of this equation
 by $\Sigma(|\bk|)$, we get
\begin{equation*}
  \eta \hat{\bk} \cdot \nabla W_{12}(\bx,\hat{\bk}) = \int_{S^{d-1}}
f(\hat{\bp}\cdot \hat{\bk} ;|\bk|) \left[e^{i\eta (\hat{\bp} - \hat{\bk})\cdot \bpsi(\bx)} W_{12}(\bx,\hat{\bp}) - W_{12}(\bx,\hat{\bk}) \right]
d\hat{\bp}.
\end{equation*}
We follow the idea used in Ref.~\onlinecite{BalRam04} and define a new
function which takes account the phase shift
\begin{equation*}
\widetilde{W}_{12} (\bx, \hat{\bk}) = e^{i\eta \hat{\bk} \cdot \bpsi(\bx)}W_{12}(\bx,\hat{\bk}).
\end{equation*}
Then it is easy to verify that $\widetilde{W}_{12}$ satisfies
\begin{equation}
\eta \hat{\bk} \cdot \nabla \widetilde{W}_{12}(\bx,\hat{\bk}) - i\eta^2 \Sigma_{\rm a}(\bx,\hat{\bk}) \widetilde{W}_{12}(\bx,\hat{\bk}) = \int_{S^{d-1}} f(\hat{\bp}\cdot \hat{\bk};|\bk|) \left[ \widetilde{W}_{12}(\bx,\hat{\bp}) - \widetilde{W}_{12}(\bx,\hat{\bk}) \right] d\hat{\bp}.
\label{eq:W12bdiff}
\end{equation}
Here, the intrinsic ``complex absorption" coefficient is given
by 
\begin{equation}
\label{eq:defabsorption}
\Sigma_{\rm a}(\bx,\hat{\bk}) =  \hat{\bk} \cdot \nabla \big( \hat{\bk} \cdot\bpsi(\bx) \big) = 
\sum_{i,j=1}^d \hat{k}_i \hat{k}_j \partial_{x_i} \psi_j(\bx)=\Tr \big( (\hat{\bk} \otimes \hat{\bk}) \nabla \bpsi(\bx) \big), 
\end{equation}
 where $\hat{\bk} \otimes \hat{\bk}$
is the projection matrix $\hat{\bk} \hat{\bk}^t$ and $\Tr$ means taking the trace.

Since $\widetilde{W}_{12}$ approximates $W_{12}$ as $\eta$ goes to zero, it suffices to consider the limit of $\widetilde{W}_{12}$. Abusing notations, we still denote this function by $W_{12}$. To start the formal derivation, we substitute the ansatz
\begin{equation}
\label{eq:DIFexp}
W_{12} = W^{(0)}_{12} + \eta W^{(1)}_{12} + \eta^2 W^{(2)}_{12} + \cdots
\end{equation}
into Eq.~\eqref{eq:W12bdiff}. Equating the terms that are of equal order in $\eta$ we  find
\begin{eqnarray}
\label{eq:diffexpa}
&\mathcal{O}(1):  &0 = \int_{S^{d-1}}f(\hat{\bp}\cdot \hat{\bk};|\bk|)  \left[W^{(0)}_{12} (\hat{\bp}) - W^{(0)}_{12}(\hat{\bk}) \right] d\hat{\bp},\\
&\mathcal{O}(\eta):  &\hat{\bk} \cdot \nabla W^{(0)}_{12} = \int_{S^{d-1}} f(\hat{\bp}\cdot \hat{\bk};|\bk|)  \left[W^{(1)}_{12}(\hat{\bp}) - W^{(1)}_{12}(\hat{\bk}) \right] d\hat{\bp},
\label{eq:diffexpb}
\\
&\mathcal{O}(\eta^2):  & \hat{\bk} \cdot \nabla W^{(1)}_{12}  - i\Sigma_{\rm a}(\bx,\hat{\bk}) W^{(0)}_{12}   = \int_{S^{d-1}} f(\hat{\bp}\cdot \hat{\bk};|\bk|) \left[W^{(2)}_{12}(\hat{\bp}) - W^{(2)}_{12}(\hat{\bk}) \right] d\hat{\bp}.
\label{eq:diffexpc}
\end{eqnarray}

To study these equations, we define the integral operator from $L^2(S^{d-1})$ to $L^2(S^{d-1})$:
\begin{equation}
{\mathcal K} h (\hat{\bk}) : = \int_{S^{d-1}} f(\hat{\bp}\cdot \hat{\bk};|\bk|) h(\hat{\bp}) d\hat{\bp}.
\end{equation}
Using this definition, Eq. \eqref{eq:diffexpa} can be recast as $({\mathcal K} - {\mathcal I}) W^{(0)} = 0$,
where ${\mathcal I}$ is the identity operator.
Note that we have $\int_{S^{d-1}} f (\hat{\bk}\cdot \hat{\bp};|\bk|)d\hat{\bp} = 1$.
We assume also that $f$ is uniformly bounded from up
and below by positive numbers. In this case, the constant function is known to be the only
eigenvector of ${\mathcal K}$ corresponding to the eigenvalue one\cite{DLv6}. Furthermore, the integral equation $({\mathcal K}-{\mathcal I})h = v$
is solvable only if $\int_{S^{d-1}} v d\hat{\bp} = 0$ (Fredholm alternative).

Using these results, Eq. \eqref{eq:diffexpa} shows that $W^{(0)}_{12}$ does not
depend on the direction variable $\hat{\bk}$, so $W^{(0)}_{12} = W^{(0)}_{12}(\bx)$. Eq.~\eqref{eq:diffexpb} 
 relates $W^{(1)}_{12}$ to $W^{(0)}_{12}$. In fact, we need to solve the integral equation
\begin{equation*}
({\mathcal K}-{\mathcal I}) W^{(1)}_{12} = \hat{\bk} \cdot \nabla W^{(0)}_{12}(\bx).
\end{equation*}
Let $h_j(\hat{\bk})$ be the unique solution whose integral over $S^{d-1}$ is zero to the equation
\begin{equation}
({\mathcal K}-{\mathcal I}) h_j(\hat{\bk}) = \hat{\mathbf{e}}_j \cdot \hat{\bk},
\label{eq:Kjeigen}
\end{equation}
where $\hat{\mathbf{e}}_j$ is the $j$-th unit vector in the orthonormal basis of $\R^d$. Indeed,
(\ref{eq:Kjeigen}) has a unique solution (up to an additive constant) by Fredholm alternative since
 $\hat{\mathbf{e}}_j \cdot \hat{\bk}$ integrates to zero on the sphere $S^{d-1}$.
We show in Appendix \ref{app:b} that the vector field $\bh(\hat{\bk}) = (h_j(\hat{\bk}))_{j=1}^d$
is given by
\begin{equation}
\bh(\hat{\bk}) = -\frac{\hat{\bk}}{1-g(|\bk|)},
\label{eq:hdef}
\end{equation}
with
\begin{equation}
\label{def:gk}
g(|\bk|) = G_d \int_{-1}^1 f(\mu;|\bk|) \mu (1-\mu^2)^{\frac{d-3}{2}} d\mu,
\end{equation}
where $G_d$ is a constant depending only on the dimension:
$$
G_d = \frac{2 \pi^{\frac{d-1}{2}}}{ \Gamma \big( \frac{d-1}{2}\big)} .
$$
It follows that 
 \begin{equation}
 \label{eq:expressW1diff}
 W^{(1)}_{12}(\bx,\hat{\bk}) = \bh(\hat{\bk})\cdot
\nabla W^{(0)}_{12}(\bx).
\end{equation}
Substituting this representation of $W^{(1)}_{12}$ into Eq.~\eqref{eq:diffexpc} and integrating the equation over
$S^{d-1}$, we find that the right-hand side
vanishes and we get
\begin{equation*}
\int_{S^{d-1}} \hat{\bk} \cdot \nabla \left[\bh (\hat{\bk}) \cdot \nabla W^{(0)}(\bx) \right] d\hat{\bk} - i\int_{S^{d-1}} \Tr\big((\hat{\bk} \otimes \hat{\bk}) \nabla \bpsi(\bx)\big) W^{(0)}(\bx) d\hat{\bk} = 0.
\end{equation*}
Carrying out these integrals, we find
\begin{equation}
-{\bf D} (|\bk|): \nabla \nabla W^{(0)}(\bx) - \frac{\varpi_d i}{d} (\nabla \cdot \bpsi(\bx)) W^{(0)}(\bx) = 0.
\label{eq:DiffLimSg1}
\end{equation}
Here, the symbol $:$ denotes the Frobenius inner product of two matrices, $d$ is
the space dimension, and $\varpi_d$ is the area of the sphere $S^{d-1}$:
$$
 \varpi_d = \frac{2 \, \pi^{\frac{d}{2}}}{ \Gamma\big( \frac{d}{2}\big)}.
$$ 
The diffusion matrix ${\bf D}(|\bk|)$ is given by
\begin{equation}
{\bf D}(|\bk|) = -\int_{S^{d-1}} \bh(\hat{\bk}) \otimes \hat{\bk} d\hat{\bk}.
\label{eq:Ddef}
\end{equation}
Substituting (\ref{eq:hdef}) into \eqref{eq:Ddef}, we find that 
\begin{equation}
{\bf D} (|\bk|)= \frac{\varpi_d}{d(1-g(|\bk|))} \Id,
\end{equation}
 where $\Id$ is the identity matrix. 
 The limiting diffusion equation for $W^{(0)}_{12}$ which is denoted by $W^{{\rm d}}_{12}$ from now on becomes
\begin{equation}
 - \nabla \cdot \frac{1}{(1-g(|\bk|))} \nabla W^{{\rm d}}_{12}(\bx) - {\color{red}i (\nabla \cdot \bpsi(\bx))} W^{{\rm d}}_{12} (\bx)= 0, \quad \bx \in X.
\label{eq:DiffLimSs}
\end{equation}

For the autocorrelation functions $W^{{\rm d}}_{11}$ and $W^{{\rm d}}_{22}$, we take $\bpsi$ above to be zero and
recover the classic limiting diffusion equation
\begin{equation}
 - \nabla \cdot \frac{1}{(1-g(|\bk|))} \nabla W^{{\rm d}}_{jj}(\bx)  = 0, \quad \bx\in X, \quad \quad j=1,2.
\label{eq:DiffLimjjSs}
\end{equation}
The above diffusion equations \eqref{eq:DiffLimSs} and \eqref{eq:DiffLimjjSs} should be equipped
with proper boundary conditions. We impose that
\begin{equation}
W^{{\rm d}}_{jl}(\bx) = q(\bx), \quad \bx \in \partial X.
\label{eq:DiffLimBC}
\end{equation}
Here, $q(\bx)$ models the incoming light intensity at the boundary. It can be derived from the
boundary condition $p(\bx,\hat{\bk})$ in \eqref{eq:RTEBC} of the RTE. If $p$ is independent of $\hat{\bk}$, then $q = p$. 
The case when $p$ is anisotropic requires a careful boundary layer multiscale analysis as developed in Refs.~\onlinecite{Chandra60, BLP79}, which shows that
 there exists a linear operator (local in $\bx$) that maps $p$ to $q$. In the three-dimensional case when $f$ is constant (isotropic scattering) and  $p(\bx, \hat{\bk}) = \tilde{p}(\bx,\mu)$ where $\mu = -\hat{\bk} \cdot \bnu(\bx)$, this map is given by
\begin{equation}
q(\bx) = \int_0^1 \tilde{p}(\bx, \mu) H(\mu) \frac{\mu}{2} d\mu,
\label{eq:ChandraHfunc}
\end{equation}
where $H(\mu)$ is the so-called Chandrasekhar $H$-function.
The evaluation of $H$ can be found, for instance, in Section 1.5 in Ref.~\onlinecite{BLP79}.
As a result the support of $q$ is spatially limited to the part $\partial X_{\rm i}$ of the boundary $\partial X$
where the laser beam is applied.

\subsection{The case of moderate shifts}
\label{sec:ms}%
In this subsection, the amplitude of the shift $\bphi$ is assumed to be moderate and of the same order as the wavelength so that 
\begin{equation}
|\bk| \bphi(\bx) = \bpsi(\bx) ,
\end{equation}
for some function $\bpsi(\bx)$ of order one.
The RTE (\ref{eq:rte12it}) for $W_{12}$ then takes the form (after dividing by $\Sigma(|\bk|)$):
\begin{equation*}
  \eta \hat{\bk} \cdot \nabla W_{12}(\bx,\hat{\bk}) = \int_{S^{d-1}}
f(\hat{\bp}\cdot \hat{\bk} ;|\bk|) \left[e^{i (\hat{\bp} - \hat{\bk})\cdot \bpsi(\bx)} W_{12}(\bx,\hat{\bp}) - W_{12}(\bx,\hat{\bk}) \right]
d\hat{\bp}.
\end{equation*}
We  define a new function which takes account the phase shift
\begin{equation*}
\widetilde{W}_{12} (\bx, \hat{\bk}) = e^{i \hat{\bk} \cdot \bpsi(\bx)}W_{12}(\bx,\hat{\bk}).
\end{equation*}
It satisfies
\begin{equation}
\eta \hat{\bk} \cdot \nabla \widetilde{W}_{12} - i\eta \Sigma_{\rm a}(\hat{\bk},\bx) \widetilde{W}_{12} = \int_{S^{d-1}} f(\hat{\bp}\cdot \hat{\bk};|\bk|) \left[ \widetilde{W}_{12}(\bx,\hat{\bp}) - \widetilde{W}_{12}(\bx,\hat{\bk}) \right] d\hat{\bp},
\label{eq:W12bdiffbis}
\end{equation}
where the intrinsic ``complex absorption" coefficient $\Sigma_{\rm a}$ is given
by (\ref{eq:defabsorption}).
We substitute the ansatz
\begin{equation}
\label{eq:DIFexpbis}
\widetilde{W}_{12} = \widetilde{W}^{(0)}_{12} + \eta \widetilde{W}^{(1)}_{12}  + \cdots
\end{equation}
into Eq.~\eqref{eq:W12bdiffbis}. Equating the terms that are of equal order in $\eta$ we  find
\begin{eqnarray}
\label{eq:diffexpabis}
&\mathcal{O}(1):  &0 = \int_{S^{d-1}}f(\hat{\bp}\cdot \hat{\bk};|\bk|)  \left[\widetilde{W}^{(0)}_{12} (\hat{\bp}) - \widetilde{W}^{(0)}_{12}(\hat{\bk}) \right] d\hat{\bp},\\
&\mathcal{O}(\eta):  &\hat{\bk} \cdot \nabla \widetilde{W}^{(0)}_{12} - i\Sigma_{\rm a}(\bx,\hat{\bk}) \widetilde{W}^{(0)}_{12} = \int_{S^{d-1}} f(\hat{\bp}\cdot \hat{\bk};|\bk|)  \left[\widetilde{W}^{(1)}_{12}(\hat{\bp}) - \widetilde{W}^{(1)}_{12}(\hat{\bk}) \right] d\hat{\bp}.
\label{eq:diffexpbbis}
\end{eqnarray}
As in the previous subsection, 
Eq. \eqref{eq:diffexpabis} shows that $\widetilde{W}^{(0)}_{12}$ does not
depend on the direction variable $\hat{\bk}$, so $\widetilde{W}^{(0)}_{12} = \widetilde{W}^{(0)}_{12}(\bx)$. Eq.~\eqref{eq:diffexpbbis} 
 relates $\widetilde{W}^{(1)}_{12}$ to $\widetilde{W}^{(0)}_{12}$. 
 By Fredholm alternative, the compatibility equation for the resolution of this equation is that the left-hand side has an integral over the sphere  equal to zero, which imposes (since the integral of $\hat{\bk}$ is zero):
 \begin{equation*}
 0 =  \int_{S^{d-1}} \Sigma_{\rm a}(\bx,\hat{\bk}) \widetilde{W}^{(0)}_{12} (\bx) d\bk = \frac{\varpi_d}{d}  (\nabla \cdot \bpsi(\bx) )  \widetilde{W}^{(0)}_{12} (\bx).
\end{equation*}
As a consequence, provided the displacement field is not divergence free (more precisely, we assume that the set of points $\bx \in X_{\rm s}$
such that $\nabla \cdot \bphi(\bx)=0$ is negligible), then $W_{12}(\bx,\hat{\bk})\approx W_{12}^{\rm d}(\bx)$ which is the solution
of 
\begin{equation}
\begin{aligned}
- \nabla \cdot \frac{1}{(1-g(|\bk|))} \nabla W^{{\rm d}}_{12}(\bx)  &= 0,  \quad & \bx \in X^c_{\mathrm{s}},\\
W^{{\rm d}}_{12}(\bx) & = 0, \quad & \bx \in X_{\mathrm{s}}.
\end{aligned}
\label{eq:DiffLimLSbis}
\end{equation}
This equation is equipped with the boundary condition \eqref{eq:DiffLimBC}. The diffusion limits for $W^{{\rm d}}_{11}$ and $W^{{\rm d}}_{22}$
 are given by \eqref{eq:DiffLimjjSs} with boundary condition \eqref{eq:DiffLimBC}.

\subsection{The case of large shifts}
\label{subsec:largeshift}%
In this subsection, we consider the case when the shift $\bphi$ of the scatterers
has large amplitude so that $|\bk||\bphi| \gg 1$. As we have derived in Section \ref{sec:rte},
the RTE for $W_{12}$ takes the form   \eqref{eq:rte12ls}. Consider the limit
$\eta = 1/\Sigma \ll 1$, these equations can be written as
\begin{equation*}
\begin{aligned}
\eta \hat{\bk} \cdot \nabla W_{12} (\bx,\hat{\bk})+ W_{12} (\bx,\hat{\bk})&= 0, \quad & (\bx,\hat{\bk}) \in X_{\mathrm{s}} \times S^{d-1}, \\
\eta \hat{\bk} \cdot \nabla W_{12} (\bx,\hat{\bk})+ W_{12} (\bx,\hat{\bk})&= \int_{S^{d-1}} f(\hat{\bk} \cdot \hat{\bp};|\bk|)  W_{12}(\bx, \hat{\bp}) d\hat{\bp}, \quad & (\bx,\hat{\bk}) \in X^c_{\mathrm{s}} \times S^{d-1}.
\end{aligned}
\end{equation*}

On the unshifted region $X^c_{\mathrm{s}}$, the second line of the RTEs above takes the classic
form and its diffusion limit is \eqref{eq:DiffLimjjSs}. On the other hand,
sending $\eta$ to zero, we deduce from the first line of the equations above that $W_{12}$ goes to zero
 in the shifted region $X_{\mathrm{s}}$ in the limit. Consequently, the diffusion limit for the
 cross-correlation $W^{{\rm d}}_{12}$ in the case of large shift is:
\begin{equation}
\begin{aligned}
- \nabla \cdot \frac{1}{(1-g(|\bk|))} \nabla W^{{\rm d}}_{12}(\bx)  &= 0,  \quad & \bx \in X^c_{\mathrm{s}},\\
W^{{\rm d}}_{12}(\bx) & = 0, \quad & \bx \in X_{\mathrm{s}},
\end{aligned}
\label{eq:DiffLimLS}
\end{equation}
and it is equipped with the boundary condition \eqref{eq:DiffLimBC}. The diffusion limits for $W^{{\rm d}}_{11}$ and $W^{{\rm d}}_{22}$
 remain unchanged; that is, they are given by \eqref{eq:DiffLimjjSs} with boundary condition \eqref{eq:DiffLimBC}.
These are the same equations as in Subsection \ref{sec:ms}.

\subsection{Speckle pattern correlation in the diffusion regime}

In this subsection, we revisit the formula \eqref{eq:C12formRTE} and rewrite it in terms of the functions
 in the diffusion limit. As before, we  denote  by $\partial X_{\rm m}$ the part of the boundary where measurement is taken. 
We naturally assume that the part $\partial X_{\rm i}$ of the boundary that is illuminated by the incident laser beam has an empty intersection with the part $\partial X_{\rm m}$ where the outgoing light is measured. 
Therefore Eq.~(\ref{eq:DiffLimBC}) implies that the leading-order term
$W^{{\rm d}}_{jl}(\bx)$ in the diffusion approximation of the direction-resolved correlation function
 $W_{jl}(\bx,\hat{\bk})$ vanishes on $\partial X_{\rm m}$. It is then necessary to look for the first-order corrections 
in the expansion of $W_{jl}(\bx, \hat{\bk})$. This is discussed in detail in Ref.~\onlinecite{rossum} and we follow the ideas there.

In the diffusion regime,  using the expansion \eqref{eq:DIFexp} with $W_{12}^{(1)}$ given by (\ref{eq:expressW1diff})
and using the similar results for $W_{11}^{(1)}$ and $W_{22}^{(1)}$, we can write $W_{jl}(\bx,\hat{\bk})$ as
\begin{equation}
\label{eq:expandeta}
W_{jl}(\bx,\hat{\bk}) = W^{{\rm d}}_{jl}(\bx) - \frac{\eta}{1-g(|\bk|)} \hat{\bk} \cdot \nabla W^{{\rm d}}_{jl}(\bx) + \cdots,
\end{equation}
where $W^{{\rm d}}_{jl}(\bx)$ is the function involved in the diffusion equations \eqref{eq:DiffLimSs} \eqref{eq:DiffLimjjSs} \eqref{eq:DiffLimLS}. 
This expansion is valid only inside the domain $X$ and not at the boundary $\partial X$, again because of the presence of 
a boundary layer which gives rise to a correction of order $\eta$ that cancels the first-order corrective term in (\ref{eq:expandeta})
for $\bx \in \partial X_{\rm m}$ and $\bnu(\bx) \cdot \hat{\bk} <0$.
If, however, we carry out the calculation with this expansion formally, then we find that 
\begin{equation*}
\int_{\hat{\bk} \cdot \bnu(\bx) > 0}  W_{jl}(\bx,\hat{\bk}) d\hat{\bk} = -\frac{\eta}{1-g(|\bk|)}  \int_{\hat{\bk} \cdot \bnu(\bx)>0}  \hat{\bk}d\hat{\bk}  \cdot \nabla W^{{\rm d}}_{jl}(\bx) = - \frac{C_d\eta}{1-g(|\bk|)} \bnu(\bx) \cdot \nabla W^{{\rm d}}_{jl}(\bx),
\end{equation*}
where $C_d$ is a constant that depends on the dimension.
The second equality follows from the decomposition $\hat{\bk}  = (\hat{\bk}\cdot \bnu(\bx)) \bnu(\bx) + \hat{\bk}_\perp$ where $\hat{\bk}_\perp$ is perpendicular
 to $\bnu(\bx)$, and the fact that the contribution of $\hat{\bk}_\perp$ averages to zero because of symmetry.
 
In fact the result obtained in this formal way is correct up to the value of the constant $C_d$.
The exact value of $C_d$ depends on the form of $f$,
it can be obtained by the multiscale analysis of the boundary layer, and it can be evaluated numerically
when $f$ is constant in particular (see Sec. X.A in Ref.~\onlinecite{rossum}).
It follows that the correlation of speckle patterns in the diffusion regime is given by
\begin{equation}
C_{12} =  \frac{\displaystyle \Big| \int_{\partial X_{\rm m}} \bnu(\bx) \cdot \nabla W^{{\rm d}}_{12}(\bx) d\bx \Big|^2}
{\displaystyle \int_{\partial X_{\rm m}} \bnu(\bx) \cdot \nabla W^{{\rm d}}_{11}(\bx)d\bx \, \int_{\partial X_{\rm m}} \bnu(\bx) \cdot \nabla W^{{\rm d}}_{22}(\bx)d\bx }.
\label{eq:C12formDIF}
\end{equation}
We remark that in the above formula, the constant ${C}_d \eta/(1-g)$
is cancelled out, so the desired result does not depend on the  value of $C_d$.

\subsection{Numerical simulations of the diffusion equation model}

In this subsection, we show some numerical simulations which
confirm that the diffusion equation model derived in this paper for the
cross correlation $W_{12}$ captures the loss of correlation in
speckle patterns. The numerical simulations are in accordance with
the experimental measurements published in Refs.~\onlinecite{Bossy07,
Bossy09}.

Let the domain $X$ be a two-dimensional square $(-1,1)\times
(-1,1)$, and let a sequence of circles $S(r_n)$ centered at
$(0,0)$ with increasing radius $r_n$ model the wave front of the
elastic wave introduced by the ultrasound modulation. Let
$C_{12}^{(n)}$ be the correlation calculated as in
\eqref{eq:C12formDIF} on the right side of the square domain $X$ 
where the two random media are those when
the wave fronts are at $S(r_n)$ and $S(r_{n-1})$ respectively.
Since $\bphi$ models the shift of the scatterers of these two
random media, the support of $\bphi$ is the union of the
supports of the elastic waves at the two instants, and it is
enclosed inside the circle $S(r_n)$. We assume that 
the shift is large enough (i.e. larger than the optical wavelength)
so that the formulas obtained in the large shift regime in Section \ref{subsec:largeshift}
are valid.

\begin{figure}[hb]
\begin{center}
\begin{minipage}{0.3\linewidth}
\begin{center}
\begin{tikzpicture}[scale=2]
\draw[style=thick] (-1,-1) rectangle (1,1);
\foreach \r in {0.1,0.3,0.5}
\draw[style=dashed,thick] (0,0) circle (\r);
\end{tikzpicture}
\newline
(a)
\end{center}
\end{minipage}
\begin{minipage}{0.3\linewidth}
\begin{center}
\begin{tikzpicture}[scale=2]
\draw[style=thick] (-1,-1) rectangle (1,1);
\path[fill=gray] (0,0) circle (0.2);
\foreach \r in {0.1,0.3,0.5}
\draw[style=dashed,thick] (0,0) circle (\r);
\end{tikzpicture}
\newline
(b)
\end{center}
\end{minipage}
\begin{minipage}{0.3\linewidth}
\begin{center}
\begin{tikzpicture}[scale=2]
\draw[style=thick] (-1,-1) rectangle (1,1);
\path[fill=gray] (0,0.1) circle (0.2);
\foreach \r in {0.1,0.3,0.5}
\draw[style=dashed,thick] (0,0) circle (\r);
\end{tikzpicture}
\newline
(c)
\end{center}
\end{minipage}
\end{center}
\caption{ \label{fig:Comp} Computational domains and elastic wave fronts. (a)
The dashed lines are the wave fronts $S(r_n)$'s. (b) An optical absorber with radius
 $r_a = 0.2$ centered at $(0,0)$. (c) An optical absorber with radius $r_a$ located at $(0,0.1)$.}
\end{figure}
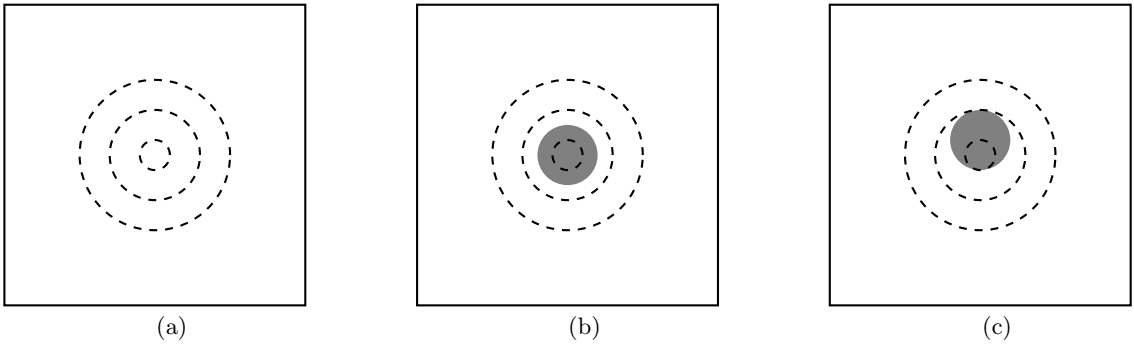

To evaluate $C_{12}^{(n)}$, we need to calculate $W_{11}, W_{22}$
and $W_{12}$. For the first two functions, we solve the diffusion
equation \eqref{eq:DiffLimjjSs} with unit diffusion coefficient on
the whole domain $X$ with boundary condition \eqref{eq:DiffLimBC},
which is taken as $q = 1$ on the left side and $q = 0$ on the
three other sides of the square domain $X$ (which models an illumination from the left). 
For $W_{12}$, since it vanishes on the support of $\bphi$ 
which has outer boundary $S(r_n)$, we solve the
first equation of \eqref{eq:DiffLimLS} on the exterior of the ball
enclosed by $S(r_n)$; the boundary condition is
\eqref{eq:DiffLimBC} on $\partial X$ and $W_{12} = 0$ on the inner
boundary $S(r_n)$. These configurations of computational domains and
elastic wave fronts are illustrated in Fig.~\ref{fig:Comp}(a).

To demonstrate the effect of optical absorbers, we also consider
the case when such an absorber with radius $r_a = 0.2$ is located at
$(0,0)$ and $(0,0.1)$ respectively, as illustrated in
Fig.~\ref{fig:Comp}(b-c). They are referred to as the case with
centered and non-centered absorbers respectively. In these cases,
the equation for $W_{11}$ and $W_{22}$ are solved outside the absorber because light is completely absorbed inside. For $W_{12}$,
the equation is solved outside the union of the absorber and the
ball enclosed the circle $S(r_n)$.

Note that for simplicity, the diffusion coefficient and the
boundary condition $q$ on the left boundary are chosen to have unit
value. This does not affect the results of the simulation. Indeed,
the diffusion constant is cancelled out in \eqref{eq:C12formDIF};
further, when the boundary condition $q$ in
\eqref{eq:ChandraHfunc} is uniform in $\bx$ on the left side, its
constant value  will be cancelled out in
\eqref{eq:C12formDIF} as well.

\begin{figure}[h]
\begin{center}
\includegraphics[height=8cm]{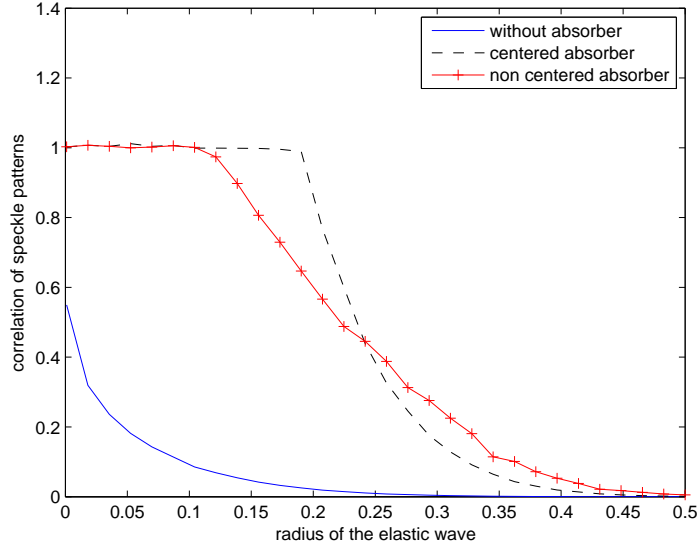}
\end{center}
\caption{\label{fig:C12} The correlation of consecutive speckle
patterns during the propagation of a circular elastic wave.}
\end{figure}

Finally we plot $C_{12}^{(n)}$ as a function of $r_n$ in
Fig.~\ref{fig:C12}. We see that when there is no optical absorber,
the correlation of speckle patterns drops immediately when elastic
wave forms. On the other hand, when optical absorber is present,
the correlation starts to decay only after the elastic wave exits
the absorber, that is at $r = 0.2$ for Fig.~\ref{fig:Comp}(b) and
at $r=0.1$ for Fig.~\ref{fig:Comp}(c). Hence the decay of
correlation is sensitive to the location of the absorber. This can
be exploited further for medical imaging purposes.

\appendix

\section{Derivation of Equation (\ref{wigner12eps})}
\label{app:A}%
We first write the equations satisfied by $u_1^\eps(\bx-\eps\by/2)$ and $\bar{u}_2^\eps(\bx+\eps \by/2)$:
\begin{eqnarray*}
\frac{\eps^2}{2} \Delta_\bx u^\eps_1 \big(\bx - \frac{\eps \by}{2} \big) &=& - \frac{1}{2} u^\eps_1  \big(\bx - \frac{\eps\by}{2}\big) 
- \sqrt{\eps} V  \big( \frac{\bx}{\eps} - \frac{\by}{2} \big) u^\eps_1  \big(\bx - \frac{\eps}{2}\by\big)   ,\\
\frac{\eps^2}{2} \Delta_\bx \bar{u}^\eps_2 \big(\bx + \frac{\eps\by}{2} \big) &=& - \frac{1}{2} \bar{u}^\eps_2  \big(\bx + \frac{\eps\by}{2}\big) 
- \sqrt{\eps} V  \big( \frac{\bx}{\eps} + \frac{\by}{2} + \bphi(\bx + \frac{\eps \by}{2}) \big) \bar{u}^\eps_2  \big(\bx + \frac{\eps\by}{2}\big) .
\end{eqnarray*}
We multiply the first equation by $\bar{u}_2^\eps(\bx+\eps \by/2)$ and the second one by $u_1^\eps(\bx-\eps\by/2)$
and we substract the two resulting equations:
\begin{eqnarray*}
&&\frac{\eps^2}{2} \Big[  \bar{u}^\eps_2 \big(\bx + \frac{\eps\by}{2} \big) \Delta_\bx u^\eps_1 \big(\bx - \frac{\eps \by}{2} \big) 
-u^\eps_1 \big(\bx - \frac{\eps \by}{2} \big)  \Delta_\bx \bar{u}^\eps_2 \big(\bx + \frac{\eps\by}{2} \big) \Big] \\
&&=  \sqrt{\eps} \Big[ V  \big( \frac{\bx}{\eps} + \frac{\by}{2} + \bphi(\bx + \frac{\eps \by}{2}) \big)- V  \big( \frac{\bx}{\eps} - \frac{\by}{2} \big) \Big]
u^\eps_1  \big(\bx - \frac{\eps}{2}\by\big)   \bar{u}^\eps_2 \big(\bx + \frac{\eps\by}{2} \big) .
\end{eqnarray*}
By multiplying by $\exp( i \bk \cdot \by) / [\eps (2\pi)^d]$ and by integrating in $\by$ we find
$$
L^\eps (\bx,\bk) = R^\eps (\bx,\bk)
$$
where we have defined 
\begin{eqnarray*}
R^\eps (\bx,\bk)
&=&  \frac{1}{\sqrt{\eps}}\int \Big[ V  \big( \frac{\bx}{\eps} + \frac{\by}{2} + \bphi(\bx + \frac{\eps \by}{2}) \big)- V  \big( \frac{\bx}{\eps} - \frac{\by}{2} \big) \Big]
u^\eps_1  \big(\bx - \frac{\eps}{2}\by\big)   \bar{u}^\eps_2 \big(\bx + \frac{\eps\by}{2} \big) \frac{e^{i \bk \cdot \by}}{(2\pi)^d} d\by ,\\
L^\eps (\bx,\bk)&=& \frac{\eps}{2} \int \Big[ \bar{u}^\eps_2 \big(\bx + \frac{\eps\by}{2} \big) \Delta_\bx u^\eps_1 \big(\bx - \frac{\eps \by}{2} \big) 
-u^\eps_1 \big(\bx - \frac{\eps \by}{2} \big)  \Delta_\bx \bar{u}^\eps_2 \big(\bx + \frac{\eps\by}{2} \big) \Big] \frac{e^{i \bk \cdot \by}}{(2\pi)^d} d\by .
\end{eqnarray*}
The right-hand side $R^\eps(\bx,\bk)$ is the one that appears in (\ref{wigner12eps}), it remains to simplify the left-hand side $L^\eps(\bx,\bk)$.
By rewriting $\Delta_\bx$ we have
\begin{eqnarray*}
L^\eps (\bx,\bk)&=& - \int \Big[\bar{u}^\eps_2 \big(\bx + \frac{\eps\by}{2} \big) \nabla_\by\cdot \nabla_\bx u^\eps_1 \big(\bx - \frac{\eps \by}{2} \big) 
+u^\eps_1 \big(\bx - \frac{\eps \by}{2} \big)  \nabla_\by \cdot \nabla_\bx \bar{u}^\eps_2 \big(\bx + \frac{\eps\by}{2} \big) \Big] \frac{e^{i \bk \cdot \by}}{(2\pi)^d} d\by .
\end{eqnarray*}
By integrating by parts in $\by$:
\begin{eqnarray*}
L^\eps (\bx,\bk)&=&  
\int 
\Big[ \nabla_\by \bar{u}^\eps_2 \big(\bx + \frac{\eps\by}{2} \big) \cdot \nabla_\bx u^\eps_1 \big(\bx - \frac{\eps \by}{2} \big) 
+ \nabla_\by u^\eps_1 \big(\bx - \frac{\eps \by}{2} \big)  \cdot \nabla_\bx \bar{u}^\eps_2 \big(\bx + \frac{\eps\by}{2} \big) \Big]  \frac{e^{i \bk \cdot \by}}{(2\pi)^d} d\by \\
&& + \int 
\Big[\bar{u}^\eps_2 \big(\bx + \frac{\eps\by}{2} \big) \nabla_\bx u^\eps_1 \big(\bx - \frac{\eps \by}{2} \big) 
+u^\eps_1 \big(\bx - \frac{\eps \by}{2} \big)  \nabla_\bx \bar{u}^\eps_2 \big(\bx + \frac{\eps\by}{2} \big) \Big] \cdot \nabla_\by \frac{e^{i \bk \cdot \by}}{(2\pi)^d} d\by \\
&=&  \frac{2}{\eps} \int 
\Big[   \nabla_\bx \bar{u}^\eps_2 \big(\bx + \frac{\eps\by}{2} \big) \cdot \nabla_\bx u^\eps_1 \big(\bx - \frac{\eps \by}{2} \big) 
- \nabla_\bx u^\eps_1 \big(\bx - \frac{\eps \by}{2} \big)  \cdot \nabla_\bx \bar{u}^\eps_2 \big(\bx + \frac{\eps\by}{2} \big) \Big]  \frac{e^{i \bk \cdot \by}}{(2\pi)^d} d\by 
\\
&& + i \bk \cdot \nabla_\bx \int 
\bar{u}^\eps_2 \big(\bx + \frac{\eps\by}{2} \big)  u^\eps_1 \big(\bx - \frac{\eps \by}{2} \big)  \frac{e^{i \bk \cdot \by}}{(2\pi)^d} d\by \\
&=& i \bk \cdot \nabla_\bx W_{12}^\eps  (\bx,\bk)  ,
\end{eqnarray*}
which is the desired result.

\section{Derivation of Equation (\ref{eq:hdef})} 
\label{app:b}%
We consider
$$
({\mathcal K}-{\mathcal I})(-\hat{\bk} \cdot \hat{\mathbf{e}}_j) = \hat{\bk} \cdot \hat{\mathbf{e}}_j - \int_{S^{d-1}} f(\hat{\bp}\cdot \hat{\bk};|\bk|) \hat{\bp} \cdot \hat{\mathbf{e}}_j d\hat{\bp} .
$$
Let ${\bf Q}_{\hat{\bk},j}$ be an orthogonal matrix so that ${\bf Q}_{\hat{\bk},j}\hat{\bk} = \hat{\mathbf{e}}_j$. 
Since $\hat{\bp} \cdot  \hat{\mathbf{e}}_j ={\bf Q}_{\hat{\bk},j}\hat{\bp} \cdot {\bf Q}_{\hat{\bk},j} 
\hat{\mathbf{e}}_j $, we have
\begin{eqnarray}
\nonumber
({\mathcal K}-{\mathcal I})(-\hat{\bk} \cdot \hat{\mathbf{e}}_j)
&= &\hat{\bk} \cdot \hat{\mathbf{e}}_j - \int_{S^{d-1}} f({\bf Q}_{\hat{\bk},j}\hat{\bp} \cdot \hat{\mathbf{e}}_j ; |\bk|) \, {\bf Q}_{\hat{\bk},j}\hat{\bp} \cdot {\bf Q}_{\hat{\bk},j} 
\hat{\mathbf{e}}_j d\hat{\bp}\\
 &=&\hat{\bk} \cdot \hat{\mathbf{e}}_j - \int_{S^{d-1}} f(\hat{\bp} \cdot \hat{\mathbf{e}}_j; |\bk|) \, \hat{\bp} \cdot {\bf Q}_{\hat{\bk},j} \hat{\mathbf{e}}_j d\hat{\bp}.
\label{eq:appb:eq2}
\end{eqnarray}
In the last equality, we changed the variable ${\bf Q}_{\hat{\bk},j} \hat{\bp}$ to
$\hat{\bp}$. We have the decomposition
\begin{equation*}
{\bf Q}_{\hat{\bk},j} \hat{\mathbf{e}}_j = [\hat{\mathbf{e}}_j \cdot {\bf Q}_{\hat{\bk},j} \hat{\mathbf{e}}_j] \hat{\mathbf{e}}_j + c \hat{\mathbf{e}}_{j\perp} = [{\bf Q}_{\hat{\bk},j} \hat{\bk} \cdot {\bf Q}_{\hat{\bk},j} \hat{\mathbf{e}}_j] \hat{\mathbf{e}}_j + c \hat{\mathbf{e}}_{j\perp} 
= (\hat{\bk} \cdot \hat{\mathbf{e}}_j) \hat{\mathbf{e}}_j + c\hat{\mathbf{e}}_{j\perp},
\end{equation*}
where $c \hat{\mathbf{e}}_{j\perp}$ is perpendicular to $\hat{\mathbf{e}}_j$. 
By symmetry,
the contribution of $c\hat{\mathbf{e}}_{j\perp}$ to the spherical integral in (\ref{eq:appb:eq2}) vanishes. We can then check that
\begin{equation*}
({\mathcal K}-{\mathcal I})(-\hat{\bk} \cdot \hat{\mathbf{e}}_j) = \hat{\bk} \cdot \hat{\mathbf{e}}_j 
- (\hat{\bk} \cdot \hat{\mathbf{e}}_j) \int_{S^{d-1}} f(\hat{\bp} \cdot \hat{\mathbf{e}}_j; |\bk|) \hat{\bp} \cdot \hat{\mathbf{e}}_j d\hat{\bp} = (1-g(|\bk|)) \, \hat{\bk} \cdot  \hat{\mathbf{e}}_j  ,
\end{equation*}
with
$$
g(|\bk|) = \int_{S^{d-1}} f(\hat{\bp}\cdot \hat{\mathbf{e}}_j ; |\bk|)  \hat{\bp}\cdot \hat{\mathbf{e}}_j d\hat{\bp} ,
$$
which does not depend on $j$ and is given by (\ref{def:gk}). Note that $g( |\bk|)<1$.
If we define $h_j(\hat{\bk})$ by
\begin{equation}
h_j(\hat{\bk}) = -\frac{\hat{\bk} \cdot \hat{\mathbf{e}}_j}{1-g(|\bk|)},
\label{eq:hjdef}
\end{equation}
then we can now check that it solves \eqref{eq:Kjeigen} and that 
it integrates to zero on $S^{d-1}$, which completes the proof.

\end{document}